\documentclass[aps,prd,reprint,nofootinbib,superscriptaddress]{revtex4-2}

\usepackage{xcolor}
\usepackage{graphicx}  
\usepackage{subfigure}
\usepackage{multirow}
\usepackage{bm}       
\usepackage{amsfonts}  
\usepackage{mathalfa}
\usepackage{amsmath}
\usepackage{mathrsfs}
\usepackage{euscript}
\usepackage{amssymb} 
\usepackage{cancel}
\usepackage{braket}
\usepackage{float}
\DeclareMathOperator\arctanh{arctanh}

\usepackage{enumitem} 
\usepackage{hyperref}
\makeatletter
\def\l@subsubsection#1#2{} 
\makeatother

\hypersetup{
   pdftitle={Fuzzy worldlines},
   colorlinks=true,
   linkcolor=black,
   urlcolor=blue
}

\begin{document}

\renewcommand{\sectionautorefname}{Section}

\title{Time-of-flight fuzziness from deformed relativistic symmetries}

\author{Giulia Gubitosi}  
\affiliation{Dipartimento di Fisica Ettore Pancini, Università di Napoli “Federico II”,
Complesso Univ.\ Monte S.\ Angelo, I-80126 Napoli, Italy}
\affiliation{INFN, Sezione di Napoli, Complesso Univ.\ Monte S.\ Angelo, I-80126 Napoli, Italy}
\author{Pietro Pellecchia}
\affiliation{Dipartimento di Fisica Ettore Pancini, Università di Napoli “Federico II”,
Complesso Univ.\ Monte S.\ Angelo, I-80126 Napoli, Italy}
\affiliation{INFN, Sezione di Napoli, Complesso Univ.\ Monte S.\ Angelo, I-80126 Napoli, Italy}
\author{Alessandro Pinto}
\affiliation{BCAM -- Basque Center for Applied Mathematics, Alameda de Mazarredo 14, E-48009 Bilbao, Bizkaia, Spain}

\begin{abstract}
A general expectation in quantum gravity is that quantum properties of spacetime manifest themselves as fuzziness, causing an irreducible uncertainty in the measurement of spacetime-related observables.
In this work, we derive the time-of-flight fuzziness for free particles within a noncommutative spacetime model with $\kappa$-Poincaré deformed relativistic symmetries. 
To retain the full quantum structure of the theory we work in  the  corresponding noncommutative space of worldlines. This provides a convenient framework for constructing quantum states and deriving the probability distributions for the parameters characterizing  particle trajectories, allowing us to obtain the probability distribution of particle  times of flight.
Our results reproduce the well-known systematic time-of-flight correction, leading to energy-dependent departures from the special-relativistic expectation, and simultaneously predict a novel stochastic contribution to the time of flight. 
The systematic contribution scales with the ratio between the particle energy and the  quantum-gravity energy scale, amplified by the propagation distance. The stochastic contribution scales as the square root of the product of the quantum-gravity length scale and the mean travel time, multiplied by a prefactor which depends on the particle velocity.
\end{abstract}

\maketitle 
%\tableofcontents

\section{Introduction}\label{sec:intro}

Quantum gravity is widely expected to profoundly affect the properties of spacetime at microscopic scales, due to the incompatibilities between its descriptions in the quantum and gravitational theories. 
Two much-studied possibilities, with wide-ranging phenomenological implications,  are spacetime fuzziness and deformed relativistic symmetries. 
Spacetime fuzziness encodes the idea that localizability is ill-defined when accounting for quantum and gravitational effects, so that spacetime observables acquire an intrinsic uncertainty~\cite{Kempf:1994su, Garay:1998wk, Kempf:1998em}. 
Deformations of relativistic symmetries have been proposed in order to accommodate the quantum gravity scale (expected to be of the order of the Planck scale) as an observer-independent quantity~\cite{Amelino-Camelia:2000cpa, Amelino-Camelia:2000stu,Kowalski-Glikman:2002iba, Magueijo:2001cr}, and find preliminary support in a number of quantum gravity theories~\cite{Matschull:1997du, Majid:1999tc, Amelino-Camelia:1999jfz, Meusburger:2003ta, Freidel:2005me, Cianfrani:2016ogm, Amelino-Camelia:2016gfx, Rosati:2017toi}.

In this work, we investigate how these two possible manifestations of quantum spacetime emerge within a common fundamental framework, focusing on the time-of-flight observable.

Time-of-flight analyses have proven to be powerful tests of departures from relativistic symmetries, exploited in the context of high-energy astrophysics~\cite{Addazi:2021xuf}. Comparing the time of flight of a low-energy particle (which is assumed to be negligibly affected by quantum-spacetime properties) to the one of a high-energy particle emitted at the same time by a given source,  the  delay between the detection time of the two particles is typically modeled as  \cite{Amelino-Camelia:1997ieq}  
\begin{equation}\label{time delay standard illustrativo}
\Delta t \simeq \Delta t_{sr}-\frac{E}{E_{QG}} \, d \, ,
\end{equation}
where $\Delta t_{sr}$ is the standard special-relativistic delay  due to possible velocity difference between the two particles, and we included the leading-order correction term, proportional to the ratio between the  energy  \(E\) of the high-energy particle and the parameter \(E_{QG}\) which sets the energy scale where quantum-gravity effects become relevant, multiplied by  the distance traveled \(d\).  

While most time-of-flight studies focus on systematic effects, i.e.\ deterministic contributions like the one $E/E_{QG}$ displayed in Eq.~\eqref{time delay standard illustrativo}, it has been suggested that fuzzy spacetime properties might produce a stochastic contribution to the time delay~\cite{Vasileiou:2015wja}.
However, whereas several studies derive systematic effects from quantum spacetime models~\cite{Amelino-Camelia:2000cpa, KowalskiGlikman:2004tz, AmelinoCamelia:2011bm, Barcaroli:2015eqe, Amelino-Camelia:2013uya, Amelino-Camelia:2011ebd}, fuzzy effects have so far been addressed primarily through phenomenological arguments~\cite{Diosi:1989hy, GACnature, AmelinoCamelia:1999pj, Ng:2000di, Christiansen:2005yg, AmelinoCamelia2025IRUVMixing}, with few notable exceptions \cite{Amelino-Camelia:2013nza, Carney:2024wnp, Freidel:2026hed}.
Our work derives the fuzzy-spacetime contribution to the time-delay effect from a fundamental quantum spacetime model.

We work within the framework of noncommutative spacetime geometry, which provides one of the few quantum-spacetime models in which both deformed relativistic symmetries and intrinsic spacetime uncertainty can be treated within a common well-defined mathematical structure.
Its relevance in quantum gravity research is motivated by the idea that nontrivial commutation relations between spacetime coordinates of the kind\footnote{Greek indices run from $0$ to $3$, while Latin indices run from $1$ to $3$. Throughout this work, we adopt units where $c = 1$.} $\left[\hat x^\mu,\hat x^\nu\right]=i\Theta^{\mu\nu}(\hat x)$, where $\Theta^{\mu\nu}(\hat x)$ is a matrix of functions of the coordinates with dimensions of a length squared, encode an uncertainty principle in the  measurement of  spacetime coordinates, $\Delta \hat x^\mu\Delta{\hat x^\nu}\geq |\langle \Theta^{\mu\nu}(\hat x) \rangle |$~\cite{Doplicher:1994tu, Szabo:2001kg}.

One of the most studied noncommutative spacetimes is the one known as  $\kappa$-Minkowski ~\cite{Majid:1994cy}, defined by the commutation relations\footnote{We keep the constant $\hbar$ explicit: noncommutativity is governed by the dimensional parameter $\hbar/E_{QG}=L_{QG}$, which sets a characteristic quantum-gravity motivated scale of length (or time)~\cite{Gubitosi:2021itz}. Note that $E_{QG}$, and therefore $L_{QG}$, might be taken to assume either a positive or negative sign throughout this manuscript.}
\begin{equation}\label{k minkowski}
[\hat{x}^0, \hat{x}^i] = -\frac{i\hbar}{E_{QG}} \hat{x}^i \, , \qquad [\hat{x}^i, \hat{x}^j] = 0 \, .
\end{equation}

The relativistic symmetries of this spacetime are described by the $\kappa$-Poincaré Hopf algebra~\cite{Lukierski:1991pn,Lukierski:1992dt}.  
In the so-called bicrossproduct basis~\cite{Majid:1994cy}, the generators \({\hat P}_\mu\) (translations) and \({\hat N}_i\) (boosts) satisfy the commutators:\footnote{While Hopf algebras are defined by additional structures besides the algebra of the generators, they are not relevant for what we are going to discuss, so we omit them.}
\begin{equation}\label{k-poincare algebra}
\begin{aligned}
     &[\hat N_i,\hat P_j]=\frac{i\hbar\,\delta_{ij}}{2E_{QG}}\left[E_{QG}^2\bigl(1-e^{-\frac{2\hat P_0}{E_{QG}}}\bigr)+\hat P^2\right]-\frac{i\hbar\,\hat P_i\hat P_j}{E_{QG}}\, ,\\
     &[{\hat P}_\mu, {\hat P}_\nu] = 0 \, , \qquad [{\hat N}_i, {\hat P}_0] = i \hbar \, \hat P_i \,,
\end{aligned}
\end{equation}
while the Casimir of the algebra is
\begin{equation}\label{operatore casimir}
  {\hat{\mathcal{C}}} = 4E_{QG}^2 \sinh^2\left(\frac{{\hat P}_0}{2E_{QG}} \right) - {\hat P^2} \,e^{\frac{{\hat P}_0}{E_{QG}}} \, .
\end{equation}
This structure provides a well-known realization of Doubly (or Deformed) Special Relativity (DSR)~\cite{Amelino-Camelia:2000stu}, and has been extensively studied in relation to time-of-flight analyses. The derivation of the induced systematic time delay effect in Eq.~\eqref{time delay standard illustrativo}, reviewed in \autoref{sec:preliminaries}, relies on a semiclassical realization of the $\kappa$-Poincaré model. We show that going beyond the semiclassical treatment, and taking into account the quantum properties of the model, the time delay \(\Delta t\) acquires indeed a fuzzy contribution.
To do so, we employ the construction described in \autoref{sec:worldlinespace} of the noncommutative space of worldlines \cite{Bal19, Bal21}, which  allows a more direct treatment of the quantum properties of particle trajectories.  
In this setting, we define the Wigner quasi-probability distribution associated to the conjugate velocity-intercept variables in the $\kappa$-space of worldlines (\autoref{sec:wigner}), which provides us with a workable framework that retains information about the intrinsic quantum uncertainties.   
From this, in \autoref{sec:timedelays}, we derive the corresponding time delay distribution, with its mean value and standard deviation, showing that the time delay \(\Delta t\) predicted by the $\kappa$-Poincaré symmetries also encodes 
an irreducible fuzzy contribution controlled by the characteristic length scale $L_{QG}$:
\begin{equation}
     \Delta t \simeq \Delta t_{sr} -\frac{E}{E_{QG}} d \, \pm \, g(v)\sqrt{ {\left|L_{QG}\right|} \, \Delta t_{sr}  } \, ,
\end{equation}
where $g(v)$ is a function of the high-energy-particle velocity, connected to the features of the fuzzy space of worldlines. Our construction shows that retaining the full noncommutative structure naturally complements the familiar deformation of the time of flight governed by the quantum-gravity energy scale with a genuinely stochastic contribution governed by the quantum-gravity length scale.
In the concluding \autoref{sec:conclusions}, we discuss the implications of such result and outline potential phenomenological applications.

\section{Systematic time delay from $\kappa$-Poincaré symmetries}\label{sec:preliminaries}

The systematic contribution to the time of flight induced by $\kappa$-Poincaré symmetries can be derived working in 1+1 dimensions. 
Upon performing a semiclassical limit on the spacetime commutators in Eqs.~\eqref{k minkowski}--\eqref{k-poincare algebra}, one obtains a deformed phase space with variables \((x^\mu, p_\mu)\) satisfying the Poisson brackets~\cite{Amelino-Camelia:2011uwb}
\begin{equation}\label{algebra DSR}
  \begin{aligned}
      &\{x^0, x^1\} = -\frac{x^1}{E_{QG}}\,, \quad \{x^0, p_0\} = -1\,, \quad \{x^1, p_1\} = 1\,, \\
      &\{x^0, p_1\} = \frac{p_1}{E_{QG}}\,, \qquad \{x^1, p_0\} = 0\,, \quad \{p_1, p_0\} = 0\,.
  \end{aligned}
\end{equation}
The generators in Eq.~\eqref{k-poincare algebra} are represented as functions on this phase space as follows:
\begin{equation}\label{rappresentazione}
   \begin{aligned}
        &N=p_1 x^0-\left(\frac{E_{QG}}{2} \left(1 - e^{-\frac{2 p_0}{E_{QG}} }\right) + \frac{p_1^2}{2E_{QG}}\right)x^1\,,\\
        &P_0=p_0\,, \qquad P_1=p_1\,,
   \end{aligned}
\end{equation}
so that they satisfy the  Poisson brackets~\cite{KowalskiGlikman:2002ft}
\begin{equation}\label{poisson algebra}
     \begin{aligned}
         &\{p_1, N\} = \frac{E_{QG}}{2} \left(1 - e^{-2 p_0 / E_{QG}}\right) - \frac{p_1^2}{2E_{QG}}\,, \\
         &\{p_0, p_1\} = 0 \,, \quad \{p_0, N\} = p_1 \,, 
     \end{aligned}
\end{equation}
that are compatible with the commutators of Eq.~\eqref{k-poincare algebra}.
Similarly, the Casimir is represented by the following function:
\begin{equation}\label{DSR casimir}
C = 4E_{QG}^2 \, \sinh^2\left(\frac{p_0}{2E_{QG}}\right) - p_1^2 e^{p_0 / E_{QG}}\,.
\end{equation}

This structure encodes a deformed relativistic kinematics for free particles, producing the time delay effect anticipated in Eq.~\eqref{time delay standard illustrativo}.
The derivation of this result is based on the Hamiltonian formalism using the Poisson brackets in Eq.~\eqref{algebra DSR} and Eq.~\eqref{poisson algebra} and the Casimir in Eq.~\eqref{DSR casimir} as Hamiltonian~\cite{KowalskiGlikman:2002ft, Amelino-Camelia:2012vlk, Amelino-Camelia:2011uwb, Amelino-Camelia:2013uya, Barcaroli:2015eqe}, as we review below.

\subsection{Worldlines in deformed special relativity}
The Hamiltonian constraint \(C = m^2\) yields the mass-shell condition for a free particle:
\begin{equation}\label{mass shell relation}
   4 E_{QG}^2 \sinh^2\left(\frac{p_0}{2E_{QG}}\right) - p_1^2 e^{\frac{p_0}{E_{QG}}} = m^2\,,
\end{equation}
such  that the rest-frame energy \(p_0(p_1 = 0)\) is $M = 2E_{QG}\, \sinh^{-1}\left(m/2E_{QG}\right)$, that coincides with \(m\) up to second order in \(E_{QG}^{-1}\) \cite{AmelinoCameliaBruno2001,Gub13}.
The corresponding coordinate velocity\footnote{Since translations have a nontrivial action on the spacetime coordinates defined by Eq.~\eqref{algebra DSR}, the coordinate velocity $v=\frac{\{x^1,C\}}{\{x^0,C\}}$ differs from the physical particle velocity $v_{phy}=\frac{\partial p_0(p_1)}{\partial p_1}$ \cite{mignemiVelocity, Mignemi:2004bc, MignemiRosati2019, Amelino-Camelia:2025ask}.} 
$v\equiv\frac{dx^1}{dx^0}=\frac{\{x^1,C\}}{\{x^0,C\}}$ and the momentum of the particle on-shell are:
\begin{equation}\label{velocity and momentum onshell}
   \begin{aligned}
        &v(p_0) =  \pm\frac{\sqrt{p_0^2 - M^2}}{p_0}\,, \\
        &p_1(p_0) = \pm \sqrt{p_0^2 - M^2}\left(1 - \frac{p_0}{2E_{QG}}\right)\,,
   \end{aligned}
\end{equation}
where the plus and minus sign correspond, respectively, to onward and backward particles and we have neglected terms of order $\mathcal{O}(1/E_{QG}^2)$, as well as contributions controlled by $M/E_{QG}$~\cite{MignemiRosati2019}. All results derived in the following maintain this level of approximation. Notice that the coordinate velocity retains its special-relativistic dependence on the particle energy, whereas the deformation is encoded in the  relation between momentum and energy.

Time and space translations of any function \(f\) on phase space are generated by the Poisson brackets with \(p_0\) and \(p_1\), respectively: \(\partial_{a_0} f = \{f, p_0\}\), \(\partial_{a_1} f = \{f, p_1\}\), where \(a_0\), \(a_1\) are the translation parameters \cite{Amelino-Camelia:2011ebd, Barcaroli:2015eqe}.
Integrating these equations yields the deformed action of finite translations on spacetime coordinates, denoted by \(T_{a_0, a_1}\).
Therefore, if an observer Alice assigns coordinates \((x_{A}^{0}, x_{A}^{1},p_{0}, p_{1})\) to a given phase-space point\footnote{Since the momentum space coordinates have vanishing Poisson brackets, see Eq.~\eqref{poisson algebra}, all relatively translated observers  assign the same momentum space coordinates to a given point in phase space. Therefore, for momentum space coordinates we can omit the subscript referring to the observer.}, the translated observer Bob  assigns it coordinates~\cite{Amelino-Camelia:2013uya}
\begin{equation}\label{deformed translation}
   \begin{aligned}
       (x_{B}^{0},x_{B}^{1},p_0,p_1) &= T_{a_0,a_1} \triangleright (x_{A}^{0},x_{A}^{1},p_0,p_1) = \\
       &=\left(x_{A}^{0} - a_0 + \frac{a_1}{E_{QG}}p_1,\, x_{A}^{1} - a_1,p_0,p_1\right).
   \end{aligned}
\end{equation}

These nontrivial transformations can be used to compare the  description of the worldline of a free particle with mass parameter $m$ given by the two observers. For the observer Alice, the worldline reads:
\begin{equation}\label{12}
w|_{A}:\quad x^{1}_A\left(x^{0}_A\right) =B + v \,x^{0}_A \,,
\end{equation}
where \(v\) is the coordinate velocity and the parameter $B$ defines the initial conditions.  
Applying the deformed translation in Eq.~\eqref{deformed translation}, the same worldline in Bob's coordinates reads
\begin{equation}\label{wl B}
\begin{aligned}
      w|_{B}:\quad x^{1}_B\left(x^{0}_B\right) &= 
      \left(B - a_1 + v \, a_0 - v \, a_1 \frac{p_1}{E_{QG}}\right) + v\, x^{0}_B \,.
\end{aligned}
\end{equation}

Comparing Eq.~\eqref{12} and Eq.~\eqref{wl B} one sees that, as in special relativity, spacetime translations merely shift the parameter defining the initial conditions without affecting the particle’s coordinate velocity. 
However, this shift contains an additional term with respect to the special-relativistic case, proportional to \( p_1/E_{QG} \), which induces a modification to the time of travel difference between particles with different energies, that we shall  derive in the following Subsection.

\subsection{Time delay in deformed special relativity}\label{subs:time delay dsr}

In the frame of a given source, the worldlines of two particles with the same mass $M$ emitted simultaneously read
\begin{equation}\label{worldlines 2}
  w|_{\textit{S}} : \quad x^1(x^0) \;\overset{L}{=}\;v_L\, x^0,\quad x^1(x^0) \;\overset{H}{=}\; v_H\, x^0 \,,
\end{equation}
where we denote the particle's velocity with \(v_{L,H}\),  $L$ and $H$ referring, respectively, to the lower and higher energy particle.  We assume that the low-energy particle is not affected by the deformation of kinematics described above (namely, we can neglect terms of the order of $\frac{p_L}{E_{QG}}$).

The detector rest frame, such that the low-energy particle arrives at the detector’s origin, is translated with respect to the emitter's frame with translation parameters  $a_0=\frac{d}{v_L},a_1=d$, where $d$ indicates the classical distance between source and detector.
In this frame, the two worldlines can be generically written as 
\begin{equation}\label{worldlines3}
   w|_{\textit{D}} : \quad  x^1(x^0)\;\overset{L}{=}\; {v}_{L} x^0 + B_{L}\,, \quad x^1(x^0) \;\overset{H}{=}\;{v}_{H}  x^0 + {B}_{H}\,,
\end{equation}
and comparing to Eq.~\eqref{wl B} one finds
\begin{equation}\label{bar values for time delay}
    {B}_L = 0\,, \qquad 
    {B}_H = d\left( \frac{v_H}{v_L} - 1 \right) - d \,v_H \frac{p_H}{E_{QG}}\,.
\end{equation}

The difference in arrival time at the detector of the two particles is defined as
\begin{equation}
    \Delta t \equiv \tau_L - \tau_H\,,
\end{equation}
where $\tau_L=- B_L/ v_L=0$ is the arrival time of the low energy particle and $\tau_H=- B_H/ v_H$ is the arrival time of the high-energy particle, see Eq.~\eqref{worldlines3}. 
Therefore, using Eq.~\eqref{bar values for time delay} one  obtains
\begin{equation}\label{Time delay DSR}
    \Delta t= \frac{B_H}{v_H}= d \left(\frac{v_H - v_L}{v_H \,v_L} \right) - \frac{p_H}{E_{QG}} d\,.
\end{equation}
Upon taking the momentum $p_H$ on shell via Eq.~\eqref{velocity and momentum onshell}, and neglecting terms of order \(M/E_{QG}\),
the  time delay in Eq.~\eqref{Time delay DSR} can be written as
\begin{equation}\label{time delay ricavato per bene}
     \Delta t \simeq \Delta t_{sr} -  \frac{E_H}{E_{QG}} d\,,
\end{equation}
up to the first order in $p_H/E_{QG}$, where $\Delta t_{sr} \equiv d \left( {\frac{v_H - v_L}{v_H \,v_L}} \right)$ is the standard time delay predicted in special relativity. 

This is the same formula we displayed  in \autoref{sec:intro}, where the time delay receives a  standard special-relativistic contribution plus a systematic correction term governed by the Planck energy. 
This derivation does not account for quantum spacetime fluctuations, as it relies on a semiclassical (Hamiltonian) framework.
Recovering these fluctuations requires retaining the full noncommutative structure of the model rather than its semiclassical realization. The remainder of this manuscript is devoted to deriving stochastic corrections to the time delay in Eq.~\eqref{time delay ricavato per bene}, accounting for the quantum properties of $\kappa$-Minkowski noncommutative spacetime. Rather than working  with spacetime coordinates, it will prove useful to work directly with coordinates of the corresponding noncommutative space of $\kappa$-worldlines, whose construction is reviewed in the following Section.

\section{From $\kappa$-Minkowski spacetime to $\kappa$-space of worldlines}\label{sec:worldlinespace} 

Following the treatment of Ref.~\cite{Bal19}, we briefly review the construction of both $\kappa$-Minkowski spacetime and the $\kappa$-space of oriented time-like worldlines as homogeneous spaces of the $\kappa$-Poincaré group in the bicrossproduct basis.

\subsection{Minkowski spacetime and space of time-like worldlines}
We start by constructing the standard (commutative) space of worldlines as the homogeneous space of the Poincaré Lie group $P(1+1)$. The corresponding Poincaré Lie algebra $\mathfrak{p}(1+1)$
with the basis $\left( P_{0},P_{1},N\right) $, consisting of the
generators of time translations, space translations and boosts, respectively,  satisfies the commutation relations:\footnote{In order to match the notation of Ref.~\cite{Bal19}, in this Subsection we use  the abstract, dimensionless, generators and group coordinates. We will reinstate physical units in the following Subsection. Note that abstract generators $ X$ are related to physical generators $\hat X$ via $ \hat X=i\hbar X$. }
\begin{equation}
 [P_{0},P_{1}]=0\,,\quad[N,P_{0}]=P_{1}\,,\quad[N,P_{1}]=P_{0}\,.
\end{equation}
As explained in detail in Refs.~\cite{Bal19,Bal21,Bal22}, Minkowski spacetime $\mathcal{M}$ and the space of time-like worldlines $\mathcal{W}$ can be defined as homogeneous spaces of the Poincaré group $P(1+1)$, obtained by quotienting it with respect to the Lie subgroups $L$ and $H$, generated by the subalgebras $\mathfrak{l}=\text{span}\{N\}$ and $\mathfrak{h}=\text{span}\{P_{0}\}$, respectively:
\begin{equation}\label{spacetime and worldline space}
    \mathcal{M}=P(1+1)/L\,,\qquad\mathcal{W}=P(1+1)/H\,.
\end{equation}

Using a faithful representation of the Poincaré algebra $\{\rho(P_0), \rho(P_1), \rho(N)\}$, convenient parameterizations of Minkowski spacetime and the space of worldlines are, respectively
\begin{equation}\label{parameterization for spacetime}
 g_{\mathcal{M}}=\exp\left[x^{0}\rho\left(P_{0}\right)\right] \exp\left[x^{1}\rho\left(P_{1}\right)\right]\exp\left[\xi\,\rho\left(N\right)\right]\,,
\end{equation}
and
\begin{equation}\label{parameterization for wl space}
     g_{\mathcal{W}}=\exp\left[\eta\,\rho\left(N\right)\right] \exp\left[y^{1}\rho\left(P_{1}\right)\right]\exp\left[y^{0}\rho\left(P_{0}\right)\right]\,.
\end{equation}
These parameterizations endow \(\mathcal{M}\) and \(\mathcal{W}\) with well-defined coordinates \((x^0,x^1)\) and \((\eta,y^1)\), respectively. 
Additionally, Eqs.~\eqref{parameterization for spacetime}-\eqref{parameterization for wl space} imply \cite{Bal21}
\begin{equation}\label{map}
    \begin{aligned}
       & x^0 = y^0 \cosh \eta + y^1 \sinh \eta\,, \\
       & x^1 = y^0 \sinh \eta + y^1 \cosh \eta\,, \\
       & \xi = \eta\,,
    \end{aligned}
\end{equation}
so that, by comparing with a generic worldline in spacetime,
\begin{equation}\label{the wl}
    x^1(x^0) = v \, x^0 + B\,,
\end{equation}
one finds that the intrinsic parameters of the worldline, namely the intercept $B$ and velocity $v$, can be written in terms of the  coordinates in the space of worldlines $(\eta,y^1)$ as~\cite{Bal21}
\begin{equation}\label{eq:Bv}
    B = \frac{y^1}{\cosh \eta}\,, \qquad v = \tanh \eta\,.
\end{equation}
This establishes a one-to-one correspondence between points in the space of worldlines \(\mathcal{W}\) and oriented time-like worldlines in \(\mathcal{M}\).

\subsection{Deformation and quantization}
Endowing the Poincaré group with the Poisson-Lie structure associated with the classical $r$-matrix $r = \frac{1}{E_{QG}}\, (N \wedge P_1)$ and projecting the corresponding Sklyanin bracket onto $\mathcal{M}$ and $\mathcal{W}$ using the parameterizations in Eqs.~\eqref{parameterization for spacetime}-\eqref{parameterization for wl space} yields Poisson structures that are covariant under the Poisson--Lie action of the Poincaré group~\cite{Bal19,Bal21,Bal22}. 
Explicitly, one finds the following Poisson structure on Minkowski spacetime
\begin{equation}\label{algebra spacetime classica}
\{ x^0, x^1 \} = -\frac{x^1}{E_{QG}}\,,
\end{equation}
while the Poisson structure of the space of worldlines is
\begin{equation}\label{algebra wl eta y classica}
\{ y^1, \eta \} = \frac{1}{E_{QG}} \left( \cosh \eta - 1 \right)\,.
\end{equation}
Then, using the change of variables in Eq.~\eqref{eq:Bv}, the Poisson bracket of the physical worldline parameters \(v\) and \(B\) reads 
\begin{equation}\label{algebra wl B v classica}
  \{ v, B \} = { -}\frac{1}{E_{QG}} \left(1 - v^2\right) \left( 1 - \sqrt{1 - v^2} \right)\,.
\end{equation}

The classical Poisson structures on $\mathcal{M}$ and $\mathcal{W}$ are promoted to  noncommutative operator algebras via a standard quantization procedure~\cite{Bal19}. 
The resulting algebras are covariant under the coaction of the $\kappa$-Poincaré quantum group in the bicrossproduct basis, Eq.~\eqref{k-poincare algebra}, consistently constructed from the same $r$-matrix.

In particular, the quantization of Eq.~\eqref{algebra spacetime classica} yields the noncommutative $\kappa$-Minkowski spacetime already described in \autoref{sec:intro}
\begin{equation}\label{indeterminazione x0 x1} 
    [\hat{x}^0,\hat{x}^1]=-\frac{i\hbar}{E_{QG}}\hat{x}^1\,.
\end{equation}
In the following, we will make use of the commutation relation arising from the quantization of the physical worldline parameters $(v,B)$,
\begin{equation}\label{indeterminazione v,B} 
    [\hat{v},\hat{B}]={ -}\frac{i\hbar}{E_{QG}}\left(1-\hat{v}^{2}\right)\left(1-\sqrt{1-\hat{v}^{2}}\right)\,.
\end{equation}
This can be interpreted as the relation between self-adjoint operators $\hat{B}$ and $\hat{v}$ on a suitable Hilbert space of worldlines $\mathcal{H_W}$ \cite{Bal21} (this Hilbert space is defined rigorously in the following Section).
For a normalized state $|\psi\rangle\in\mathcal{H_W}$, one can thus compute the standard deviation over $|\psi\rangle$ of a given self-adjoint operator $\hat{a}$ as $D_\psi a\equiv\sqrt{\langle\hat{a}^2\rangle_{\psi}-\langle\hat a\rangle_{\psi}^2}$, where $\langle\cdot\rangle_{\psi}\equiv\langle\psi|\cdot|\psi\rangle$ denotes expectation values over $|\psi\rangle$.
In this framework, the commutator in Eq.~\eqref{indeterminazione v,B} yields the uncertainty relation
\begin{equation}\label{vera indeterminazione}
    D_\psi v\,D_\psi B\;\ge\;\frac{{\left|L_{QG}\right|}}{2}\,
   \left| \left\langle\left(1-\hat v^{2}\right)\left(1-\sqrt{1-\hat v^{2}}\right)\right\rangle_{\psi}\right|\,,
\end{equation}
which characterizes the intrinsic uncertainty in particle worldlines. 

In closing this Section, we note that $L_{QG}=\frac{\hbar}{E_{QG}}$ governs noncommutativity of both spacetime and worldlines coordinates, Eqs.~\eqref{indeterminazione x0 x1}-\eqref{indeterminazione v,B}, and the associated minimum uncertainty relations, such as Eq.~\eqref{vera indeterminazione} (for spacetime coordinates, this has been studied, \textit{e.g.}, in Ref.~\cite{Lizzi:2019wto}). If one takes the limit $\hbar\to 0$, while keeping $E_{QG}$ finite, the uncertainty relations become trivial, and the commutators in Eqs.~\eqref{indeterminazione x0 x1}-\eqref{indeterminazione v,B} reduce to the corresponding Poisson brackets defined in Eqs.~\eqref{algebra spacetime classica}-\eqref{algebra wl B v classica}. Therefore, this limiting procedure leads to the semiclassical framework which was used in \autoref{sec:preliminaries} to derive the systematic contribution to the time delay \eqref{time delay ricavato per bene}. 
The further limit $ E_{QG}\to \infty$ reproduces classical Minkowski spacetime and standard special relativity~\cite{AmelinoCamelia2010}.
Indeed, as already mentioned, the systematic contribution to the time delay emerges within the deformed-special-relativity frameworks~\cite{Amelino-Camelia:2000cpa, KowalskiGlikman:2004tz, AmelinoCamelia:2011bm}, which are thought of as possible quantum-gravity manifestations in a regime where both the Planck constant $\hbar$ and Newton's constant $G$ are negligible ($G \to 0$, $\hbar \to 0$) but with finite ratio, so that the Planck length vanishes while the Planck energy remains finite~\cite{Arzano:2022har}.\footnote{As previously mentioned, the energy scale $E_{QG}$ and the length scale $L_{QG}$ amount to, respectively, the Planck energy $E_p \equiv \sqrt{\hbar/G}$ and the   Planck length $L_p \equiv \sqrt{\hbar G}$ up to numerical factors of order one.}

In order to derive the  time delay contribution encoding not only  deformations governed by the quantum-gravity energy scale  but also ‘quantization’ effects governed by the quantum-gravity length scale, we stay within the  regime in which $\hbar$ is  finite. 
In the next Section, we therefore start from the uncertainty relation in Eq.~\eqref{vera indeterminazione} between $\hat{B}$ and $\hat{v}$, allowing us to study the probabilistic nature of the velocity and intercept variables.

\section{Probability distribution of fuzzy worldlines}\label{sec:wigner}

Having established that worldline parameters become noncommuting operators, we now construct their quantum states and the corresponding probability distributions. This allows us to describe particle trajectories probabilistically and provides the statistical framework needed to derive time-of-flight fluctuations in the next Section. 

The velocity and intercept operators $(\hat v, \hat B)$ satisfy a nonlinear commutation relation, Eq.~\eqref{indeterminazione v,B}, which, upon defining the uncertainty function 
\begin{equation}\label{funzione di indeterminazione}
    \gamma(v)=\left(1-v^{2}\right)\left(1-\sqrt{1-v^{2}}\right)\,,
\end{equation}
can be written as 
\begin{equation}\label{commutatione semplice}
    \left[\hat B,\hat v\right]=i L_{QG} \, \gamma(\hat v)\,.
\end{equation}

In this Section, we apply the construction developed in Appendix \ref{app:generalwigner} to associate quantum states to Eq.~\eqref{commutatione semplice}, defined in an appropriate Hilbert space.
Equipped with this tool, we investigate the minimal fuzzy effects that can be ascribed to quantum uncertainty by  specializing to generalized coherent states, which provide the closest quantum counterpart of classical worldlines, as they (approximately) saturate the uncertainty principle, Eq.~\eqref{vera indeterminazione}. 
We analyze the statistical properties of the observables $(\hat v, \hat B)$ within the Wigner quasiprobability formalism \cite{Wig32} which provides a phase-space description in terms of classical distributions that faithfully encode the statistical properties of the corresponding quantum states, and has been previously applied to the quantum space of worldlines in \cite{Bal21}. In this framework, the quantum properties of worldlines implied by the commutation relation  Eq.~\eqref{commutatione semplice} are encoded in probability distributions for the velocity and intercept variables.

\subsection{Hilbert space of worldlines and coherent states}
The operators $\hat v$ and $\hat B$ in Eq.~\eqref{commutatione semplice}
can be defined   as self-adjoint operators on a suitable Hilbert space through the following representation on eigenstates of $\hat v$:
\begin{equation}\label{rappre}
    \begin{aligned}
    \hat{v}\,\psi(v) &= v\,\psi(v)\,,\\
    \hat{B}\,\psi(v) &=
     i L_{QG}\!\left[\gamma(v)\partial_v
    + \frac12 \partial_v\gamma(v)\right]\psi(v)\,,\\
    \hat{\gamma}\,\psi(v) &= \gamma(v)\,\psi(v)\,.
\end{aligned}
\end{equation}
For later convenience, we also define the auxiliary function $\Gamma(v)$ such that $d\Gamma=dv/\gamma(v)$, that is
\begin{equation}
     \Gamma(v)= \arctanh v -   \frac{1+\sqrt{1-v^{2}}}{v}\,.
\end{equation}

Since $\gamma(v)$ is even in its domain, we may, without loss of generality, restrict the analysis from the full velocity domain $(-1,1)$ to $(0,1)$, therefore focusing on forward-directed worldlines. 
Representation in Eq.~\eqref{rappre} then ensures that $\hat v$ and $\hat B$ are well-defined self-adjoint operators acting on the Hilbert space $\mathcal{H_W}\equiv L^2([0,1])$.

In this domain, $\Gamma(v)$ and $\gamma(v)$ satisfy all the properties that are required to develop a well-defined generalization of coherent states associated to Eq.~\eqref{commutatione semplice}, as discussed in 
 Appendix \ref{app:generalwigner}. 
In the velocity representation, a coherent state on the Hilbert space of worldlines $\mathcal{H_W}$ reads
\begin{equation}\label{STATO COERENTE}
    \chi_{\bar{B}, \bar{\Gamma}}(v) =\frac{\exp\left[ - \frac{\big( \Gamma(v) - \bar\Gamma\big)^2}{2\Omega^2 \left|L_{QG}\right|}  - i \frac{\bar{B}}{|L_{QG}|} \Gamma(v) \right]}{ \left( \pi \Omega^2 \left|L_{QG}\right|\gamma(v)^2 \right)^{\frac{1}{4}}}  \, ,
\end{equation}
where  we have defined the  expectation values $\bar{B} \equiv \langle \hat{B} \rangle_\chi$ and $\bar \Gamma \equiv \langle \Gamma(\hat{v}) \rangle_\chi$. The parameter $\Omega^2$ has dimensions of an inverse length and  controls the spread of the coherent state, in close analogy with the width of Gaussian coherent states in ordinary quantum mechanics.
As for coherent states in standard quantum mechanics,  $\bar \Gamma$ and $\bar B$ correspond to their classical values.\footnote{Note that the classical values are affected by the deformed symmetries described by the Poisson brackets in Eqs.~\eqref{algebra DSR} and~\eqref{poisson algebra}.} 
The classical value of $\Gamma(v)$ is given by evaluating the function  on the classical value of the velocity, $\bar v$, so that $\bar\Gamma=\Gamma(\bar v)$. Due to the nonlinear relation between $\Gamma$ and $v$, the expectation value of the velocity, however, is only equal to its classical value up to factors of order $L_{QG}$, as shown below.

As anticipated at the beginning of this Section and demonstrated in  Appendix \ref{app:generalwigner}, we note that the state $ \chi_{ \bar{B}, \bar{\Gamma}}(v)$ saturates the uncertainty principle, Eq.~\eqref{vera indeterminazione}, up to first order in $L_{QG}$, that is 
\begin{equation}\label{incertezza}
     D_\chi v\,D_\chi B\;\simeq\;\frac{|L_{QG}|}{2}
    \left\langle \gamma(\hat v) \right\rangle_{\chi}\,.
\end{equation}

\subsection{Wigner distribution and localizability}
The Wigner quasiprobability distribution associated to the coherent state defined in Eq.~\eqref{STATO COERENTE} is 
\begin{equation}\label{Wigner function new}
    W(v, B) = \frac{ \exp \left[ -\frac{\left( \Gamma(v) - \Gamma(\bar{v}) \right)^{2}}{\Omega^{2} \left|L_{QG}\right|} - \frac{\Omega^{2}}{\left|L_{QG}\right|} (B - \bar{B})^{2} \right] }{\pi \left|L_{QG}\right| \, \gamma(v)}\, ,
\end{equation}
as shown in Appendix \ref{app:generalwigner}. 
This function encodes the statistical properties of the $\kappa$-space of worldlines: velocity $v$ and intercept $B$ are no longer sharp, but take values spread around the classical reference $(\bar{v},\bar{B})$ according to the distribution $W(v,B)$.
Note that, as in standard quantum mechanics, the Wigner function of a coherent state is a proper probability distribution: it is not only real and normalized, but also non-negative for each allowed value of $v$ and $B$. 

The associated marginal distributions for the intercept and the velocity, computed by integrating on the other variable, are
\begin{equation}\label{marginali nuove}
   \begin{aligned}
    W_{B}(B) &= \sqrt{\frac{\Omega^2}{\pi\left|L_{QG}\right|}} \exp \left[ -\frac{\Omega^2}{\left|L_{QG}\right|} (B - \bar{B})^2 \right], \\[2mm]
      W_{v}(v)  &= \frac{1}{\sqrt{\pi \left|L_{QG}\right| \, \gamma(v)^2 \, \Omega^2}} \exp \left[ -\frac{(\Gamma(v) - \Gamma(\bar{v}))^2}{\Omega^2 \left|L_{QG}\right|} \right]\,.
\end{aligned}
\end{equation}
Note that the full Wigner function in Eq.~\eqref{Wigner function new} is simply the product of the two marginal distributions, $W(v,B)=W_v(v)\,W_B(B)$, just as for coherent states in ordinary quantum mechanics.
Since $W(v,B)$ is factorizable in the two variables, the covariance of $v$ and $B$ vanishes.

While the intercept distribution is a simple Gaussian, the velocity distribution is more involved.
However, the corresponding mean value and variance can be computed via Taylor expansion, as shown in Appendix \ref{app:generalwigner}:
\begin{equation}\label{medie e varianze B,v chiave nuove}
    \begin{aligned}
        E[B] &= \bar{B}\,,  \qquad  E[v] \simeq \bar{v} + \left|L_{QG}\right| \frac{\Omega^2}{4}\gamma(\bar{v}) \gamma'(\bar{v})  \,, \\
       \textit{Var}[B] &= \frac{\left|L_{QG}\right|}{2\,\Omega^2} \,,  \quad \textit{Var}[v] \simeq \frac{\left|L_{QG}\right|}{2}  \gamma(\bar{v})^2 \Omega^2\,,
    \end{aligned}
\end{equation}
where in the second line we have neglected higher order terms starting from $\mathcal{O}(L_{QG}^2)$. As we mentioned in the previous Subsection, while the expectation value of $B$ corresponds to its classical value, the expectation value of the velocity $v$ receives corrections of  order  $L_{QG}$, reflecting the nonlinear relation between $v$ and the canonical variable $\Gamma$.  These corrections induce only subleading contributions to the mean time delay between the travel times of particles with different energy, adding to the correction already induced at the semiclassical level by the deformation of symmetries, Eq.~\eqref{time delay ricavato per bene}. These  contributions are negligible with respect to the stochastic fluctuations in the time delay  induced by the variances in Eqs.~\eqref{medie e varianze B,v chiave nuove},  whose leading behavior scales as  $\sqrt{L_{QG}}$, see next Section.  

The statistical moments of Eq.~\eqref{medie e varianze B,v chiave nuove} depend on the parameter $\Omega^2$, which governs the width of the generalized coherent states, as discussed in the previous Subsection.
It can be either treated as an independent physical parameter of the model, to be constrained experimentally, or it can be linked to the other physical parameters, namely the velocity $\bar v$ (dimensionless) and the intercept $\bar B$ (with dimensions of length). In the latter case, dimensional analysis leads us to write
\begin{equation}\label{omega}
    \Omega^2=\frac{1}{|\bar{B} \, f(\bar{v})|}\,,
\end{equation}
where  $f(\bar{v})$ is a generic function of $\bar v$. 
This relation induces the expected limiting behaviors of the intercept and velocity, and is compatible  with the well-known localizability properties of $\kappa$-Minkowski spacetime, as we show in the following.

Using the ansatz~\eqref{omega}, we can rewrite Eq.~\eqref{medie e varianze B,v chiave nuove} as
\begin{equation}\label{final statistical moments}
    \begin{aligned}
        E[B] &= \bar{B}\,,  \qquad E[v] \simeq \bar{v} +  \left|L_{QG}\right|\frac{\gamma(\bar{v}) \gamma'(\bar{v})}{4\,|\bar{B} \,f(\bar{v})|} \,, \\
       \textit{Var}[B]& = \frac{|L_{QG}|}{2} |\bar{B}\,f(\bar{v})| \,,   \quad \textit{Var}[v] \simeq \frac{|L_{QG}|}{2}  \frac{\gamma(\bar{v})^2 }{|\bar{B}\,f(\bar{v})|}\,.
    \end{aligned}
\end{equation}

The intercept variance $\textit{Var}[B]$ in Eq.~\eqref{final statistical moments} increases as the classical distance of the worldline from the spacetime origin, given by $|\bar{B}|$, increases, while it vanishes in the limit $|\bar{B}|\to 0$, which holds for classical worldlines passing through the spacetime origin.
This can be seen as another manifestation of relative locality~\cite{AmelinoCamelia:2011bm, Amelino-Camelia:2012vlk}, known to arise in semiclassical models with $\kappa$-Poincaré symmetries~\cite{Gub13,Amelino-Camelia:2011uwb}, in  $\kappa$-Minkowski noncommutative spacetime~\cite{Amelino-Camelia:2011uwb, Lizzi:2019wto} and recently also discussed within the framework of the $\kappa$-space of worldlines itself~\cite{Bal21}.
Also note that, accounting for the ansatz in Eq.~\eqref{omega}, in the limit $|\bar{B}|\to 0$ the marginal $W_B(B)$ becomes a Dirac delta centered in zero, while $W_v(v)$ approaches a vanishing constant value.
In other words, when $|\bar{B}|\to 0$ the intercept is maximally localized while the velocity is maximally spread.

Having established the probabilistic description of quantum worldlines, we can now  determine how their intrinsic  fluctuations affect physical observables. In the next Section we derive the resulting probability distribution for the time delay.

\section{Effects of fuzziness on time delay}\label{sec:timedelays}

As illustrated in \autoref{sec:preliminaries}, models based on $\kappa$-Poincaré symmetries produce systematic modifications to particle propagation, governed by the inverse of the quantum-gravity energy scale $E_{QG}$, see Eq.~\eqref{Time delay DSR}.
Such systematic contributions are derived within a particular limit of $\kappa$-Minkowski spacetime in which the Planck constant (and hence the quantum-gravity length scale $L_{QG}$) vanishes while the energy scale $E_{QG}$ is held finite, see discussion in \autoref{sec:worldlinespace}. 

If  the  length scale $L_{QG}$ is kept finite, so that the noncommutative properties of the $\kappa$-Minkowski spacetime are fully accounted for, one can derive the ensuing fuzzy properties of particles' worldlines, as we did in the previous Section. 
Through the marginal distributions in Eq.~\eqref{marginali nuove}, we are now able to quantify fluctuations in particle trajectories affecting time delays.

\subsection{Time delay distribution}
Let us consider the same physical scenario described in \autoref{subs:time delay dsr}.
Two particles with equal mass $M$ are emitted simultaneously by a given source. As discussed at the end of the previous Section, in the source frame the parameter $\bar B$ can be known with arbitrary precision, so it is meaningful to state that the two particles are emitted in the  spacetime origin of the source frame. 
As in \autoref{subs:time delay dsr}, the detector frame is identified by  requiring that the low-energy particle passes through its spacetime origin. This condition can be imposed sharply, as the variance of the intercept parameter of a worldline, Eq.~\eqref{final statistical moments}, vanishes when the expectation value is zero.

In this setup, the  time delay between the arrival of  the high-energy particle and  the low-energy particle at the detector  has the same functional dependence on the intercept $B_H$ and velocity $v_H$ of the high-energy particle as in the deformed but commutative case, Eq.~\eqref{Time delay DSR}:
\begin{equation}\label{time delay classico}
    \Delta t=\frac{B_H}{v_H}\,.
\end{equation}
In the noncommutative scenario we are now exploring, however, $B_H$ and  $v_H$ are no longer sharp quantities. Rather, they follow the Wigner quasiprobability distribution of Eq.~\eqref{Wigner function new} centered around the values $\bar{v}_H,\bar{B}_H$. These values are  given by the sharp, but deformed, values that can be deduced from Eq.~\eqref{velocity and momentum onshell} and Eqs.~\eqref{wl B}-\eqref{bar values for time delay}: 
\begin{equation}\label{barvbarB_new}
    \bar v_H=\frac{\sqrt{E_H^2-M^2}}{E_H} \, , \quad \bar B_H= d\left( \frac{\bar v_H}{\bar v_L} - 1 \right) - d \,\bar v_H \frac{E_H}{E_{QG}} \,,
\end{equation}
where  $E_H$ is the energy of the high-energy particle, $\bar v_L=\frac{\sqrt{E_L^2-M^2}}{E_L}$ is the velocity of the low-energy particle whose energy is $E_L$, and $d$ is the classical distance between source and observer, identified by the low-energy particle.

Therefore, the time delay itself follows a probability distribution, given by the ratio of two distributed quantities. 
This distribution  can be computed via the tools of standard probability theory \cite{Ben05}, and reads
\begin{equation}\label{time d distribution nuova}
      \mathcal{T}(\Delta t)=\int_{0}^{1}dv\,v \, W(v,v\,\Delta t)\,,
\end{equation}
where $W$ is the Wigner function defined in Eq.~\eqref{Wigner function new} with central values given by Eq.~\eqref{barvbarB_new}.

The full distribution $\mathcal{T}(\Delta t)$ can be evaluated  numerically. 
For our purposes, however, it is sufficient to consider its mean value and standard deviation, which can be expressed in terms of the statistical moments of intercept and velocity derived in the previous Section.

\subsection{Mean value and standard deviation}

Using Eq.~\eqref{medie e varianze B,v chiave nuove}, a Taylor expansion of the distribution for the time delay of Eq.~\eqref{time d distribution nuova} around the central values $\bar{v}_H,\bar{B}_H$ yields the mean and standard deviation:
\begin{equation}\label{mean e std time delay generic}
    \begin{aligned}
      E[\Delta t] & 
     = \frac{\bar{B}_H}{\bar{v}_H}+\mathcal{O}\left({L_{QG}}\right)\,,\\
      D[\Delta t] & = \sqrt{\frac{\left|L_{QG}\right|}{2\,\bar{v}_H^{2}}
      \left(\frac{1}{\Omega^{2}}
      +\frac{\bar{B}_H^{2}\,\gamma(\bar{v}_H)^{2}\,\Omega^{2}}{\bar{v}_H^{2}}\right)}
      +\mathcal{O}({L_{QG}^{3/2}})\,,
    \end{aligned}
\end{equation}
where, since the leading contribution scales as ${\left|L_{QG}\right|}^{1/2}$, higher-order corrections starting from order $L_{QG}$ have been neglected.

The leading term of the mean value does not have contributions due to worldlines fuzziness, and coincides with the time delay computed in the deformed symmetry scenario, given by Eq.~\eqref{Time delay DSR} or, equivalently, Eq.~\eqref{time delay ricavato per bene}. Nontrivial effects due to fuzziness appear in the standard deviation, which scales as the square root of the length scale $L_{QG}$. 
In other words, the mean reproduces the deformed-symmetry prediction, whereas the leading manifestation of quantum fuzziness appears exclusively in the variance.

Accounting for the dimensional constraint of Eq.~\eqref{omega}, the leading term of the standard deviation reads
\begin{equation}\label{mean e std time delay nuovi}
     D[\Delta t]  = {g(\bar v_H)} \sqrt{\frac{\left|L_{QG}\right||\bar{B}_H|}{\bar{v}_H}}\,,
\end{equation}
where $g(v)$ is a dimensionless function of velocity defined as
\begin{equation}\label{g function}
    g(v)\equiv\sqrt{\frac{  v^2 f(v)^2+\gamma (v)^2}{2 v^3 f(v)}} \, .
\end{equation}
Substituting Eq.~\eqref{barvbarB_new} in Eq.~\eqref{mean e std time delay nuovi} one finds that the time delay between the two particles has mean value and standard deviation
\begin{equation}\label{mean e std time delay g}
    \begin{aligned}
      E[\Delta t] & = \Delta t_{sr} -  \frac{E_H}{E_{QG}} d\,,\\
      D[\Delta t] & = {g(\bar v_H)}\sqrt{ {\left|L_{QG}\right|} \, \Delta t_{sr} } \, ,
     \end{aligned}
\end{equation}
up to higher orders in $E_{QG}^{-1}$ and $L_{QG}$.\footnote{We remind the reader that  $d$ is the distance between source and detector identified by the low-energy particle, $\Delta t_{sr}= d \left( \frac{v_H - v_L}{v_H \,v_L}\right)$ is the standard time delay due to the velocity difference between the high-energy and low-energy particle, $E_H$ is the energy of the high-energy particle and $\bar v_H$ the associated velocity.}

Interestingly, Eq.~\eqref{mean e std time delay generic} indicates an irreducible uncertainty of the time delay: varying the expression with respect to the free parameter $\Omega^2$, one finds the minimal value
\begin{equation}\label{irreducible}
    D[\Delta t]=\sqrt{\frac{\left|L_{QG}\right|\,|\bar{B}_H|\,\gamma(\bar{v}_H)}{\bar{v}_H^{3}}} \, ,
\end{equation}
obtained when the width parameter is:
\begin{equation}
    \Omega^2=\frac{\bar{v}}{\gamma(\bar{v})\, |\bar{B}|} \, .
\end{equation}
Note that this expression is compatible with the constraint in Eq.~\eqref{omega} obtained from dimensional considerations, and the two coincide for $f(v)=\gamma(v)/v$. Additionally, with this choice of $\Omega^2$, the two variables $v$ and $B$ have equal relative variance, see also Appendix \ref{app:generalwigner}. In fact, in this case the statistical moments in Eq.~\eqref{medie e varianze B,v chiave nuove} are rewritten as 
\begin{equation}\label{final statistical moments symmetric}
    \begin{aligned}
        E[B] &= \bar{B}\,, &   \textit{Var}[B]& = \frac{|L_{QG}|}{2} \frac{|\bar{B}| \gamma(\bar v)}{ \bar v} \,, \\
       E[v] &\simeq \bar{v} +  \left|L_{QG}\right|\frac{\bar v \, \gamma'(\bar{v})}{4|\bar{B}| } \,, &  
        \textit{Var}[v]& \simeq \frac{|L_{QG}|}{2}  \frac{ \bar v \, \gamma(\bar{v})}{|\bar{B}|}\,.
    \end{aligned}
\end{equation}

Substituting Eq.~\eqref{barvbarB_new} in Eq.~\eqref{irreducible} one finds that the time delay between the two particles has mean value and minimum standard deviation fully determined by the particle mass and energy through $\bar v_H$:
\begin{equation}\label{mean e std time delay minimal}
    \begin{aligned}
      E[\Delta t] & = \Delta t_{sr} -  \frac{E_H}{E_{QG}} d\,,\\
      D[\Delta t] & = \frac{\sqrt{\gamma(\bar v_H)}}{\bar v_H}\sqrt{\left|L_{QG}\right|  \, \Delta t_{sr} } \,,
     \end{aligned}
\end{equation}
up to higher orders in $E_{QG}^{-1}$ and $L_{QG}$.

Written in this form, both Eqs.~\eqref{mean e std time delay g} and~\eqref{mean e std time delay minimal} make explicit that the quantum-gravity energy scale and the quantum-gravity length scale govern physically distinct manifestations of the same underlying noncommutative geometry: the former controls the systematic deformation of the mean propagation time, whereas the latter controls its intrinsic stochastic fluctuations.

\subsection{Phenomenological implications}
The stochastic contribution of both Eqs.~\eqref{mean e std time delay g} and \eqref{mean e std time delay minimal} exhibits a square-root dependence on the quantum-gravity length scale:
\begin{equation}
D[\Delta t]\propto \sqrt{|L_{QG}|\Delta t_{sr}}\,,
\end{equation}
analogous to the random-walk behavior previously proposed in Refs.~\cite{Diosi:1989hy, GACnature, AmelinoCamelia:1999pj,  Ng:2000di, Christiansen:2005yg}, based on phenomenological considerations and on the analysis of distance measurement procedures that take into account the interplay between the quantum properties and mass of the measurement devices exchanging light signals. In the present case, however, this scaling is derived rigorously from a fundamental noncommutative spacetime model. 
Moreover, our result predicts a nontrivial velocity-dependent prefactor, which in the minimal-uncertainty case is ${\sqrt{\gamma(\bar v_H)}}/{\bar v_H}$,  qualitatively distinguishing this framework from earlier proposals.

A first consequence of this prefactor is that the stochastic contribution becomes strongly suppressed in the ultra-relativistic regime. Indeed, since 
\begin{equation}
\lim_{v\to1} \frac{\sqrt{\gamma(v)}}{v}=0\,,
\end{equation}
the variance vanishes for particles approaching the speed of light. This behavior has important implications for experimental searches. In particular, standard astrophysical time-of-flight analyses~\cite{Addazi:2021xuf} as well as gravity-wave interferometers studies~\cite{AmelinoCamelia:1999pj} are not expected to provide the optimal setting to probe the fuzziness predicted by the $\kappa$-space of worldlines, if we expect continuity between time-like and light-like worldlines features.\footnote{A detailed description of the $v=1$ regime would require to revise our analysis starting from the $\kappa$-space of light-like worldlines, which appears as the natural follow up of this work, that we defer to future investigations.}

The stochastic contribution might instead be relevant in the low-velocity regime. Indeed, in this case the prefactor tends to a constant of order one:
\begin{equation}
\lim_{v\to0} \frac{\sqrt{\gamma(v)}}{v}=\frac{1}{\sqrt{2}}\,,
\end{equation}
in units where $c=1$.
This suggests that slow massive particles may provide a more promising arena for phenomenology.
In particular, matter-wave interferometry with atoms or molecules appears naturally suited to investigate the fluctuations described by Eq.~\eqref{mean e std time delay minimal}.  For instance, assuming that the distance \(D\) traveled by a particle of velocity \(v\) fluctuates with the same scaling of its time-of-flight, the corresponding standard deviation reads 
\begin{equation}
    \sigma_{D}=\frac{\sqrt{\gamma(v/c)}}{v/c} \sqrt{ {\left|L_{QG}\right|}\, D} \, ,
\end{equation}
where we have restored ordinary units. 
To illustrate the magnitude of the effect, let us consider a rubidium atom with mass \(M\simeq10^{-25}\,\mathrm{kg}\), moving at velocity \( v=10\,\mathrm{m/s}\) through an interferometer arm of length \(D=10\,\mathrm{m}\). Assuming a Planckian value for the characteristic length scale, \(L_{QG}\sim10^{-35}\,\mathrm{m}\), one finds a phase fluctuation 
\begin{equation}
\Delta\phi = \frac{M v\, \sigma_D}{\hbar}\sim10^{-7} \mathrm{rad} \,.
\end{equation}
Current matter-wave interferometers typically achieve phase sensitivities in the range $10^{-3}$--$10^{-4}\,\mathrm{rad}$, and some rubidium-based devices are aiming at a few \(10^{-6}\,\mathrm{rad}\) \cite{MorelYaoCladeEtAl2020}, suggesting that the fluctuations predicted in this framework could become experimentally accessible in the near future.

\section{Conclusions}\label{sec:conclusions}

In this work we have investigated the interplay between deformations of relativistic symmetries, governed by the quantum-gravity energy scale $E_{QG}$, and spacetime quantum fuzziness, governed by the quantum-gravity length scale $L_{QG}$. Starting from the $\kappa$-Poincaré Hopf algebra and the associated $\kappa$-Minkowski spacetime, we used the recently developed framework of the noncommutative space of worldlines to provide a quantum description of particle trajectories and study the joint effects of symmetry deformations and fuzziness on the time-of-flight observable.  

Within the noncommutative $\kappa$-space of worldlines, velocity and intercept of a worldline satisfy a nonlinear commutation relation, from which we defined generalized coherent states and the associated Wigner quasiprobability distributions. In this way, worldlines acquire an intrinsic probabilistic character, encoded in the probability distributions for their defining parameters.
Studying the mean and variance of these parameters, we are able to show that the statistical properties of worldlines provide a further manifestation of relative locality:  the variance of the intercept depends linearly on its expectation value and vanishes for worldlines crossing the spacetime origin, implying that distant events are intrinsically less localized than local ones. 

Our construction demonstrates how Planck-energy symmetry deformations and Planck-length quantum effects emerge from different sectors of the same underlying model. The limit $L_{QG}\to0$ at finite $E_{QG}$ reproduces the semiclassical DSR kinematical framework and the usual systematic effects associated with $\kappa$-Poincaré symmetries, while the subsequent limit $E_{QG}\to\infty$ yields ordinary special relativity. 

These features are most clearly illustrated by the 
time-of-arrival delay between particles with different energies. Indeed,  we showed that the mean value of the  time delay coincides, at leading order, with the standard prediction obtained in the deformed but commutative framework, which contains a correction with respect to the special-relativistic contribution $\Delta t_{sr}$ scaling as $\frac{E}{E_{QG}}$. Quantum fuzziness does not introduce a phenomenologically relevant shift to this mean value. Instead, it generates a genuinely new effect in the form of an irreducible stochastic contribution, such that the standard deviation of the time delay scales as $\sqrt{{|L_{QG}|\Delta t_{sr}}}$, see Eqs.~\eqref{mean e std time delay g}-\eqref{mean e std time delay minimal}.

For the specific class of coherent states considered in this work,  the result can be summarized  as
\begin{equation}
     \Delta t \simeq \Delta t_{sr}-\frac{E_H}{E_{QG}}d\,\pm\,  g(\bar v_H)\sqrt{|L_{ QG}| \Delta t_{sr} }\, ,
\end{equation}
where the second term reproduces the well-known DSR correction, while the third term represents a genuinely quantum contribution associated with the fuzziness of worldlines. The two effects are controlled by different fundamental scales and exhibit different scaling behaviors. The free function $g(v)$ can be fixed by asking that the time-delay variance is minimized, thus yielding $g(v)=\sqrt{\gamma(v)}/v$, see Eq.~\eqref{mean e std time delay minimal}.

Our results provide, to our knowledge, the first derivation of a stochastic contribution to the time of flight directly from a fundamental noncommutative spacetime model. In particular, the square-root dependence on the quantum-gravity length scale resembles the random-walk behavior previously proposed on phenomenological grounds \cite{Diosi:1989hy, GACnature, AmelinoCamelia:1999pj,  Ng:2000di, Christiansen:2005yg}, while the presence of the velocity-dependent factor $g(v)$ distinguishes our result from these earlier heuristic models.

Our analysis has been restricted to time-like worldlines. A proper treatment of massless particles would require the use of the $\kappa$-space of lightlike worldlines, constructed in Refs.~\cite{Ballesteros:2022nif, Gutierrez-Sagredo:2024jiw}. 
This appears as a natural continuation of the present work.

Finally, the framework developed here is not limited to time-of-flight observables and provides a general tool to investigate quantum fluctuations of particle trajectories. In particular, as discussed at the end of the previous Section, slow massive particles and matter-wave interferometry may offer a more promising arena to probe these effects than conventional astrophysical time-delay measurements. We leave a detailed phenomenological analysis of these possibilities to future work.

\begin{acknowledgments}
P.P.\ acknowledges helpful discussions with Domenico Frattulillo, Francesco G. Capone, Tommaso Moretti, and Federico Greco, as well as support from the WOST, WithOut SpaceTime project, supported by Grant ID 63683 from the John Templeton Foundation (JTF). 
G.G.\ thanks Ángel Ballesteros for useful comments.
A.P.\ acknowledges support by the Basque Government through the BERC 2022-2025 program, as well as the following grant: Grant PID2024-156184NB-I00 funded by MICIU/AEI/10.13039/501100011033 and cofunded by the European Union.
This work falls within the scope of the COST Action CA23130 ``Bridging high and low energies in search of quantum gravity''.
\end{acknowledgments}

\appendix
\section{Generalized coherent states}\label{app:generalwigner}
In this Appendix we derive a generalized notion of coherent states that is valid for a broad class of noncommutative spaces, controlled by 
\begin{equation}\label{commutator}
   [\hat{\beta},\hat{\alpha}]= i\,L_{QG}\,\hat \gamma \,,
\end{equation}
where  $L_{QG}$ is a real parameter with dimensions of length (this is the parameter that is identified with the quantum-gravity scale in the main body of this work), and such that  the following properties hold: 
\begin{enumerate}[label=(\alph*)] 
    \item \label{it:gamma_op}
    The operator $\hat \gamma$ can be written as $\hat{\gamma}=\gamma(\hat \alpha)$.

    \item \label{it:gamma_positive}
    The function $\gamma(\alpha)$ is real, non-negative and of type $\mathcal{C}^1$ on its domain $\mathcal{D}_\alpha\equiv[\alpha_{1},\alpha_{2}]\subseteq\mathbb{R}$.

    \item \label{it:gamma_boundary}
    The function $\gamma(\alpha)$ vanishes on the boundary: $\gamma(\alpha_1)=\gamma(\alpha_2)=0$.

    \item \label{it:Gamma_def}
    The function $\Gamma(\alpha)$, defined by the relation 
    $ d\Gamma(\alpha)= \frac{d\alpha}{\gamma(\alpha)}$, is invertible in $\mathcal{D}_\alpha$.

    \item \label{it:Gamma_range}
    The function $\Gamma(\alpha)$ spans the whole real line and diverges at the boundary: $\Gamma(\alpha_1)=-\infty$ and $\Gamma(\alpha_2)=+\infty$.
\end{enumerate}

\subsubsection*{Hilbert space and deformed uncertainty principle}
To begin with, we need a representation of this algebra such that $\hat\alpha$, $\hat{\beta}$ and $\hat{\gamma}$ are self-adjoint operators on a suitable Hilbert space.
This is achieved by requiring that the operators act  on eigenfunctions of $\alpha$  as
\begin{equation}\label{rappresentazioen}
    \begin{aligned}
    \hat{\alpha}\psi(\alpha) &= \alpha\,\psi(\alpha)\,,\\
    \hat{\beta}\psi(\alpha) &=
     i L_{QG}\!\left[\gamma(\alpha)\partial_\alpha
    + \frac12 \partial_\alpha\gamma(\alpha)\right]\psi(\alpha)\,,\\
    \hat{\gamma}\psi(\alpha) &= \gamma(\alpha)\psi(\alpha)\,,
\end{aligned}
\end{equation}
where we have used property~\ref{it:gamma_op}, which implies $[\hat \alpha,\hat \gamma]=0$, so that the two operators can be diagonalized simultaneously.
While $\hat\gamma$ and $\hat \alpha$ are trivially self-adjoint, the additional term in $\partial_\alpha \gamma$ is necessary to enforce the symmetry of $\hat \beta$, so that also $\hat \beta$ is self-adjoint under the usual scalar product  $\langle\phi|\psi\rangle\equiv\int_{\mathcal{D}_\alpha} d\alpha \, \phi^\star(\alpha)\, \psi(\alpha)$. 
In fact, after integration by parts and using Eq.~\eqref{rappresentazioen} in combination with property~\ref{it:gamma_boundary}, one can show that $\langle \psi , \hat{\beta} \psi \rangle = \langle \hat{\beta} \psi , \psi \rangle$.
Finally, Eq.~\eqref{rappresentazioen} ensures that $\hat\alpha$, $\hat{\beta}$, $\hat{\gamma}$ are self-adjoint on the Hilbert space $\mathcal H = L^2(\mathcal D_\alpha)$.

On this Hilbert space, Eq.~\eqref{commutator} induces the following uncertainty relation between $\hat\alpha$ and $\hat\beta$:
\begin{equation}\label{heisenberg}
    D_\psi \alpha\,D_\psi \beta
    \ge \frac{|L_{QG}|}{2}\left|\langle\psi|\gamma(\hat{\alpha})|\psi\rangle\right|\,,
\end{equation}
where, for any self-adjoint operator $\hat A$, the associated standard deviation is defined as
\begin{equation}
    D_\psi A
    \equiv \sqrt{\langle\psi|\hat A^2|\psi\rangle
    -\langle\psi|\hat A|\psi\rangle^2}\,.
\end{equation}

The commutator between $\hat \alpha$ and $\hat \beta$ is a deformation of the usual Heisenberg relation between conjugate operators, requiring a suitable generalization of the usual construction of coherent states, discussed in the following.

\subsubsection*{Eigenstates and scalar product}
Given the representation in Eq.~\eqref{rappresentazioen}, one can show that eigenstates of $\hat{\beta}$, such that $\hat \beta \, \varphi_\beta(\alpha)=\beta \, \varphi_\beta(\alpha)$, are given by
\begin{equation}\label{eigenstate}
    \varphi_\beta(\alpha)
    =\frac{\mathcal{N}}{\sqrt{\gamma(\alpha)}}
    \exp\!\left[-\frac{i}{|L_{QG}|}\beta\,\Gamma(\alpha)\right]\,,
\end{equation}
where $\beta\in \mathbb{R}$ and $\mathcal{N}$ is a normalization factor.

For each couple of eigenstates $ \varphi_{\beta_1}(\alpha)$ and $ \varphi_{\beta_2}(\alpha)$, the scalar product gives
\begin{equation}
    \begin{aligned}
    \langle\varphi_{\beta_1}|\varphi_{\beta_2}\rangle
    &\equiv\int_{\mathcal{D}_{\alpha}} d\alpha \, \varphi^\star_{\beta_1}(\alpha)\, \varphi_{\beta_2}(\alpha)
    =\\
    &=\mathcal{N}^2\int_{\alpha_1}^{\alpha_2} \frac{d\alpha}{\gamma(\alpha)}
      \exp\!\left[\frac{i}{|L_{QG}|}({\beta_1}-{\beta_2})\Gamma(\alpha)\right]= \\
    &=\mathcal{N}^2\int_{-\infty}^{+\infty} d\Gamma\,
      \exp\!\left[\frac{i}{|L_{QG}|}({\beta_1}-{\beta_2})\Gamma\right]
      =\\
      &=\mathcal{N}^2 \,  \left| L_{QG}\right|\,  2\pi\,\delta({\beta_1}-{\beta_2})\,,
\end{aligned}
\end{equation}
where the second line exploits the change of variables $\alpha\rightarrow\Gamma(\alpha)$ and the property~\ref{it:Gamma_range}.
Therefore, the eigenstates defined in Eq.~\eqref{eigenstate} are orthonormal upon choosing $\mathcal{N}^2\equiv \frac{1}{2\pi\left| L_{QG}\right|}$.
 
Having defined eigenstates of $\hat \alpha$ and of $\hat \beta$, any state $|\psi\rangle\in\mathcal H$ can be represented in either basis,
$\psi(\alpha)=\langle\alpha|\psi\rangle$ and
$\tilde\psi(\beta)=\langle\beta|\psi\rangle$,
related by 
\begin{equation}\label{furier}
    \begin{aligned}
    \tilde{\psi}(\beta)
    &=\frac{1}{\sqrt{2\pi |L_{QG}|}}
      \int_{\mathcal D_\alpha}
      \frac{d\alpha}{\sqrt{\gamma(\alpha)}}
      \exp\!\left[\frac{i}{|L_{QG}|}\beta\,\Gamma(\alpha)\right]\psi(\alpha)\,,\\
    \psi(\alpha)
    &=\frac{1}{\sqrt{2\pi |L_{QG}|}}
      \int_{\mathbb{R}} 
      \frac{d\beta}{\sqrt{\gamma(\alpha)}}
      \exp\!\left[-\frac{i}{|L_{QG}|}\beta\,\Gamma(\alpha)\right]\tilde{\psi}(\beta)\,.
\end{aligned}
\end{equation}

Note that the standard quantum-mechanical case is recovered for $\gamma(\alpha)=1$.
In this limit, by identifying $(\hat{\beta},\hat{\alpha})\rightarrow(\hat{x},\hat{p})$ and $L_{QG}\rightarrow\hbar$, the construction in Eq.~\eqref{furier} reduces to the usual Fourier transform between position and momentum representations in quantum mechanics. Our construction is also in agreement with results obtained in Refs.~\cite{Lizzi:2019wto, Lizzi:2018qaf} using the Mellin transform. In fact, considering the (1+1)-dimensional $\kappa$-Minkowski spacetime where 
\begin{equation}
    \left[\hat{x}^{0},\hat{r}\right]=iL_{QG}\,\hat{r}\,,
\end{equation}
and defining the function $\Gamma$ as
\begin{equation}
    \begin{aligned}
    \Gamma(r)	 &= \log(r)\qquad  \,\,\,\,\,r>0\,,\\
    \Gamma(r)	 &= \log(|r|)\qquad \,\,r<0\,,
\end{aligned}
\end{equation}
one finds that the transform in Eq.~\eqref{furier} is the Mellin transform
\begin{equation}
    \begin{aligned}
\tilde{\psi}(x^{0})	&=\dfrac{1}{\sqrt{2\pi|L_{QG}|}}\int_{0}^{\infty}dr\,\exp\left[\frac{i}{|L_{QG}|}\log(r)x^{0}\right]\psi(r)=\\
	&=\dfrac{1}{\sqrt{2\pi|L_{QG}|}}\int_{0}^{\infty}dr\,r^{\frac{ix^{0}}{|L_{QG}|}}\psi(r)=\\
	&=\dfrac{1}{\sqrt{2\pi|L_{QG}|}}\,\mathcal{M}\left\{ \psi\right\} \left(\frac{ix^{0}}{|L_{QG}|}+1\right)\,.
\end{aligned}
\end{equation}
The generalized coherent states we derive in the following Subsection are also consistent with the coherent states defined in Refs.~\cite{Lizzi:2019wto, Lizzi:2018qaf}.

\subsubsection*{Ladder operators and generalized coherent states}
To define generalized coherent states, it is convenient to work with ladder operators, as in standard quantum mechanics.
Since $\hat \beta$ and $\hat \alpha$ are not properly conjugate, we seek an operator that is canonically conjugated to $\hat \beta$.
This is the operator $\hat\Gamma\equiv\Gamma(\hat\alpha)$, where $\Gamma(\alpha)$ is the function defined in property \ref{it:Gamma_def} and
\begin{equation}\label{newcommutazione}
        [\hat{\beta},\hat{\Gamma}] = i L_{QG} \, ,
\end{equation}
as can be checked by using Eq.~\eqref{rappresentazioen}.
Ladder operators are then defined in the standard way as 
\begin{equation}\label{ladder}
     \begin{aligned}
         \hat{a}& =\frac{1}{\sqrt{2|L_{QG}|}}\left(\text{sign}(L_{QG})\,\Omega\,\hat{\beta}+\frac{i}{\Omega}\,\hat{\Gamma}\right) \, , \\
         \hat{a}^\dagger &=\frac{1}{\sqrt{2|L_{QG}|}}\left(\text{sign}(L_{QG}) \, \Omega\,\hat{\beta}-\frac{i}{\Omega}\,\hat{\Gamma}\right) \, ,
     \end{aligned}
\end{equation}
where $\Omega$ is a dimensional parameter ensuring that $\hat a$ and $\hat a^\dagger$ are dimensionless.\footnote{Note that in standard quantum mechanics $\Omega$ is in general  provided by the dynamics. For example, for a particle of mass $m$ in a harmonic-oscillator potential with frequency $\omega$ one would  have $\Omega=\sqrt{m\omega}$. For free particles, one applies a well-established procedure when defining coherent states for continuous-spectrum systems (see, \textit{e.g.}, \cite{delaTorre:2010qvr,Guerrero_2011,Maamache2016AnalyzingGC}), where $\Omega$ physically dictates the initial width of the coherent wavepacket.}

Since these ladder operators satisfy the usual relation $[\hat{a},\hat{a}^\dagger] = 1$, as can be checked by use of Eq.~\eqref{newcommutazione}, 
 a coherent state can be  defined in the usual way as the eigenstate of $\hat a$.
To do so, one needs  to solve the eigenvalues equation
\begin{equation}\label{definition coherent}
    \hat a \,\chi_{_{\delta}}(\alpha) =\delta  \, \chi_{_{\delta}}(\alpha) \, ,
\end{equation}
where $\delta$ is the displacement (complex) parameter, typically decomposed as $\delta=\delta_R+i\delta_I$. 
Using the representation of Eq.~\eqref{rappresentazioen} and the definition in Eq.~\eqref{ladder}, one can separate variables to rewrite Eq.~\eqref{definition coherent}  as 
\begin{equation}\label{eq:diff_coherent}
    \frac{\chi_{_\delta}'(\alpha)}{\chi_{_\delta}(\alpha)} = -\frac{1}{2}\frac{\gamma'(\alpha)}{\gamma(\alpha)} - \frac{1}{\Omega^2 |L_{QG}|}\frac{\Gamma(\alpha)}{\gamma(\alpha)} - i \delta\color{black} \sqrt{\frac{2}{\Omega^2 |L_{QG}|}} \frac{1}{\gamma(\alpha)} \, ,
\end{equation}
 where the prime denotes derivatives with respect to $\alpha$. This can be  integrated to find
\begin{equation}\label{risultato intermedio}
     \chi_{_{\delta}}(\alpha) = \frac{{ C \exp\left[\!-\!\frac{\big( \Gamma(\alpha) - \delta_I  \sqrt{2\Omega^2 |L_{QG}|}\big)^2}{2\Omega^2 |L_{QG}|}\!-\!\frac{i \, \delta_R \sqrt{2}}{\sqrt{\Omega^2 |L_{QG}|}}   \Gamma(\alpha) \right] }}{\sqrt{\gamma(\alpha)}}   \, ,
\end{equation}
where $C$ is an integration constant.
The mean values of $\hat \Gamma$ and $\hat \beta$ on this state can be computed directly and give
\begin{equation}\label{eq:valori_medi}
   \begin{aligned}
       &\bar\Gamma \equiv \langle \hat{\Gamma} \rangle_\chi = \delta_I  \sqrt{2\Omega^2 |L_{QG}|}\, , \\
   &\bar\beta \equiv \langle \hat{\beta} \rangle_\chi =\delta_R  \sqrt{\frac{2 |L_{QG}|}{\Omega^2}} \, ,
   \end{aligned}
\end{equation}
so that the coherent state of Eq.~\eqref{risultato intermedio} can be written as 
\begin{equation}\label{eq:stato_coerente_esplicito}
    \chi_{\bar{\beta}, \bar{\Gamma}}(\alpha) = \frac{ \exp\left[ - \frac{\big( \Gamma(\alpha) - \bar\Gamma \,\big)^2}{2\Omega^2 |L_{QG}|}  -  \frac{i}{|L_{QG}|} \bar{\beta}\,\Gamma(\alpha) \right]}{\left( \pi \Omega^2 |L_{QG}|\gamma(\alpha)^2\right)^{\frac{1}{4}}}  \, ,
\end{equation}
where we found the integration constant $C$ by requiring normalization, and we traded the subscript $\delta$ for $(\bar{\beta}, \bar{\Gamma})$, given the relation in Eq.~\eqref{eq:valori_medi} between displacement parameter and mean values.

Eq.~\eqref{eq:stato_coerente_esplicito} is a coherent state in the $\alpha$ representation.  Eq.~\eqref{furier}, allows us to write it in the $\beta$ representation:
\begin{equation}\label{coerente in beta}
    \tilde{\chi}_{\bar{\beta}, \bar{\Gamma}}(\beta) =  \frac{ \exp\left[ - \frac{\Omega^2}{2 |L_{QG}|} (\beta - \bar{\beta})^2 +  \frac{i}{|L_{QG}|} \bar\Gamma\, \beta \right] }{\left(\pi\,\Omega^{-2} |L_{QG}|\right)^{\frac{1}{4}}} \, .
\end{equation}
We note that $ \tilde{\chi}_{\bar{\beta}, \bar{\Gamma}}(\beta) $ and  $ \chi_{\bar{\beta}, \bar{\Gamma}}(\alpha)$ are the deformed version of the ordinary coherent states written in the position and momentum representation, respectively.
In fact, the substitutions $(\hat{\beta},\hat{\Gamma})\rightarrow(\hat{x},\hat{p})$, $L_{QG}\rightarrow\hbar$ and $\Omega\rightarrow\sqrt{m\omega}$ lead to the familiar coherent states of quantum mechanics: 
\begin{equation}
    \begin{aligned}
    \chi_{\bar x,\bar p}(p)&=\left(\frac{1}{\pi \hbar m\omega}\right)^{1/4}\exp\!\left[-\frac{(p - \bar p)^2}{2\hbar m \omega}-\frac{i}{\hbar}\bar x\, p\right] \, , \\
       \tilde\chi_{\bar x,\bar p}(x)&=\left(\frac{m\omega}{\pi \hbar}\right)^{1/4}\exp\!\left[-\frac{m\omega}{2\hbar}(x - \bar x)^2+\frac{i}{\hbar}\,x\, \bar p\, \right]\, .
\end{aligned}
\end{equation}

\subsubsection*{Probability distributions from coherent states}
Using the coherent state in the two representations, Eqs.~\eqref{eq:stato_coerente_esplicito} and Eq.~\eqref{coerente in beta}, one can compute the corresponding probability distributions for $\alpha$ and $\beta$:
\begin{equation}\label{modulo quadro a e b}
    \begin{aligned}
        &\left|\chi_{\bar{\beta}, \bar{\Gamma}}(\alpha)\right|^2=
        \dfrac{1}{\sqrt{\pi |L_{QG}|\, \gamma(\alpha)^2\, \Omega^2}} \, \exp\!\left[-\dfrac{\left(\Gamma(\alpha)-\bar\Gamma\right)^{2}}{\Omega^{2}|L_{QG}|}\right],\\[1mm]
          &\left|\tilde\chi_{\bar{\beta}, \bar{\Gamma}}(\beta)\right|^2=
         \sqrt{\dfrac{ \Omega^2}{\pi |L_{QG}|}} \exp\!\left[-\dfrac{\Omega^{2}}{|L_{QG}|}\left(\beta-\bar{\beta}\right)^{2}\right].
\end{aligned}
\end{equation}
The variable $\beta$ follows a Gaussian distribution with mean value and variance given by
\begin{equation}\label{media di b}
    E[\beta]=\bar \beta\, , \qquad \mathrm{Var}[\beta]=\frac{|L_{QG}|}{2 \,\Omega^{2}}\, ,
\end{equation}
while the statistical properties of $\alpha$ are more involved, due to the presence of $\gamma(\alpha)$ and $\Gamma(\alpha)$. 
However, using the properties~\ref{it:Gamma_def} and~\ref{it:Gamma_range}, one can perform the change of variable $\alpha\rightarrow\Gamma$ to rewrite the mean value of $\alpha$ as 
\begin{align}
      E[\alpha] &=\int^{+\infty}_{-\infty} d\Gamma\,\dfrac{\alpha(\Gamma)}{\sqrt{\pi \Omega^{2}|L_{QG}|}}\exp\!\left[-\dfrac{(\Gamma-\bar{\Gamma})^{2}}{\Omega^{2}|L_{QG}|}\right],
\end{align}
where $\alpha(\Gamma)$ denotes the inverse function of $\Gamma(\alpha)$.
By Taylor expanding  the integrand around $\bar{\Gamma}$, one obtains a series of Gaussian integrals that can be computed explicitly:
\begin{equation}\label{media di a}
    \begin{aligned}
    E[\alpha]\simeq & \,\alpha(\bar{\Gamma})+\frac{\Omega^{2}}{4}|L_{QG}|\,\alpha''(\bar{\Gamma})+\mathcal{O}(L_{QG}^2)=\\
    =& \, \bar{\alpha}+\frac{\Omega^{2}}{4}|L_{QG}|\,\gamma(\bar{\alpha})\,\gamma'(\bar{\alpha})+\mathcal{O}(L_{QG}^2)\, ,
\end{aligned}
\end{equation}
where we have defined $\bar\alpha\equiv\alpha(\bar \Gamma)$ and we have neglected second order terms in $L_{QG}$.
The same technique can be applied to the variance of $\alpha$ to find
\begin{equation}\label{varianza di a}
    \mathrm{Var}[\alpha]\simeq \frac{\Omega^{2}}{2}\gamma(\bar{\alpha})^{2}\,|L_{QG}|+\mathcal{O}(L_{QG}^2)\,.
\end{equation}

Standard coherent states are known to saturate the uncertainty principle.
In this deformed case, the generalized coherent state defined in Eq.~\eqref{eq:stato_coerente_esplicito} saturates the uncertainty principle, Eq.~\eqref{heisenberg}, only up to first order in $L_{QG}$, as can be seen by
using Eqs.~\eqref{media di b}--\eqref{varianza di a} to compute the product of standard deviations 
\begin{equation}
    D_\alpha\,D_\beta\simeq
        \frac{|L_{QG}|}{2}\,\gamma(\bar{\alpha})+\mathcal{O}(L_{QG}^2)\,,
\end{equation}
and the mean value of $\hat\gamma$ on the state:
\begin{equation}
     \begin{aligned}
         \langle\chi&|\hat{\gamma}|\chi\rangle=\int_{\mathcal{D}_{\alpha'}} d\alpha'  \left|\chi_{\bar{\beta},\bar{\Gamma}}(\alpha')\right|^2\! \gamma(\alpha')\,\simeq \gamma(\bar{\alpha})+\mathcal{O}(L_{QG})\,.
     \end{aligned}
\end{equation}

\subsubsection*{Wigner distribution}
We are now in a position to define a notion of joint probability distribution of $\alpha$ and $\beta$, following the construction of  the Wigner quasiprobability distribution used in quantum mechanics \cite{Wig32}.
In the deformed case we are considering, the Wigner function associated to a given state $\psi(\alpha)$ is found observing that $\beta, \Gamma$ are canonical variables, see Eq.~\eqref{newcommutazione}:
\begin{equation}
     W_\psi(\alpha,\beta)=\int_{\mathbb{R}} dq\,\,\tilde{\psi}^{*}(\beta+q)\,\tilde{\psi}(\beta-q)\,\frac{\exp\!\left[\frac{2iq}{|L_{QG}|}\,\Gamma(\alpha)\right]}{\pi |L_{QG} |\gamma(\alpha)}\, .
\end{equation}
This, for the generalized coherent states $\chi_{\bar{\beta},\bar{\Gamma}}(\alpha )$ defined in the previous Subsection, takes the  form
\begin{equation}
     W_\chi(\alpha,\beta)=\dfrac{\exp\!\left[-\dfrac{\left(\Gamma(\alpha)-\Gamma(\bar{\alpha})\right)^{2}}{\Omega^{2}|L_{QG}|}-\dfrac{\Omega^{2}}{|L_{QG}|}\left(\beta-\bar{\beta}\right)^{2}\right]}{\pi |L_{QG}|\,\gamma(\alpha)}\, .
\end{equation}
As expected for a Wigner function, its marginal distributions 
\begin{equation}
   \begin{aligned}
        W_\alpha(\alpha)&= \int_{\mathbb{R}}d\beta \,\, W_\chi(\alpha,\beta)\, ,   \\
     W_\beta(\beta)&=  \int_{\mathcal{D}_\alpha}\!d\alpha \,\, W_\chi(\alpha,\beta)\, ,
   \end{aligned}
\end{equation}
reproduce  the distributions obtained from the square modulus of $ \chi_{\bar{\beta},\bar{\Gamma}}(\alpha)$ and $ \tilde{\chi}_{\bar{\beta},\bar{\Gamma}}(\beta)$ computed in Eq.~\eqref{modulo quadro a e b}.

The Wigner quasiprobability is useful as it allows to study the statistical properties of any observable defined as a function $z=z(\alpha,\beta)$ of the variables $\alpha$ and $\beta$. 
The distribution of $z$ on a state $\psi$ is defined as
\begin{equation}
    Z(z)=\int_{\mathcal{D}_{\alpha}} d\alpha\,\left|\frac{d}{dz}f(z,\alpha)\right| \, W\!\left(\alpha,f(z,\alpha)\right),
\end{equation}
where $f(z,\alpha)$ denotes the inverse of $z(\alpha,\beta)$ with respect to $\beta$.

To summarize, we have shown that the non-standard commutator of Eq.~\eqref{commutator} allows for the definition of generalized coherent states in the Hilbert space $\mathcal{H}=L^2(\mathcal{D}_\alpha)$,  that  saturate the uncertainty relation at leading order. 
The associated Wigner function, along with the Taylor expansions for the statistical moments of any function $z(\alpha,\beta)$, provide the necessary framework to evaluate the mean and variance of the time delay discussed in the main text.

\subsubsection*{Symmetrized coherent states}
To conclude, we note that the definition of ladder operators, Eq.~\eqref{ladder}, led us to introduce the dimensional parameter $\Omega^2$.
This is analogous to the construction of coherent states of free particles in standard quantum mechanics.
In fact, in this case creation and annihilation operators are not tied to the dynamics (i.e.\ to a specific Hamiltonian), but rather to the kinematical structure of phase space, namely its commutation relations. 
By construction, a ladder operator combines a canonical coordinate and its conjugate momentum into a single complex variable. 
For such a linear combination to be well-defined, the two quantities must be made dimensionally compatible by introducing a characteristic scale.\footnote{In the case of the harmonic oscillator, the Hamiltonian lifts the degeneracy of phase space and naturally provides this scale through the combination $m\omega$, thereby selecting a preferred vacuum state. By contrast, for a free particle no such scale is dynamically selected, as the Hamiltonian preserves the full symmetry of phase space. Consequently, there is no distinguished vacuum, and all Gaussian wave packets are kinematically equivalent \cite{delaTorre:2010qvr,Guerrero_2011,Maamache2016AnalyzingGC}.}

The parameter $\Omega^2$ encodes the intrinsic widths of the coherent states. 
Indeed, the state $\chi_{\bar{\beta}, \bar{\Gamma}}$ defined in Eq.~\eqref{eq:stato_coerente_esplicito} is actually a one-parameter family of coherent states, labeled by $\Omega^2$. 
On these states, the mean values and standard deviations of the conjugate observables are
\begin{equation}\label{app mean values e standard deviations}
      \begin{aligned}
          &\bar\beta= \, \delta_R  \sqrt{\frac{2 |L_{QG}|}{\Omega^2}} \, , \qquad  \bar\Gamma =\delta_I \sqrt{2\Omega^2 |L_{QG}|}\, , \\
          &D_\beta = \,\sqrt{\frac{|L_{QG}|}{2 \, \Omega^2}} \, ,   \qquad \,  D_\Gamma = \sqrt{\frac{|L_{QG}| \, \Omega^2}{2}}   \, ,
      \end{aligned}
\end{equation}
while those of the original variable $\alpha$, already computed in Eqs.~\eqref{media di a}-\eqref{varianza di a}, read
\begin{equation}\label{app mean values e standard deviations v}
       \bar\alpha \simeq \alpha(\bar\Gamma)\, , \qquad  
       D_\alpha \simeq \sqrt{\frac{\Omega^{2}\gamma(\bar{\alpha})^{2}\,|L_{QG}|}{2}}   \, .
\end{equation}
As shown earlier, each state in this coherent state family exactly saturates the uncertainty principle in $(\beta,\Gamma)$, while it saturates the uncertainty principle in $(\alpha,\beta)$ up to first order in $L_{QG}$. 
Nevertheless, without fixing $\Omega^2$ coherent states cannot be uniquely defined, since the individual uncertainties of $\hat{\beta}$ and $\hat{\Gamma}$ would remain undetermined \cite{delaTorre:2010qvr}.

This parameter can be fully determined by choosing a particular coherent state with symmetric relative fluctuations. 
A first option is to impose that the symmetry exists between the symplectic variables $(\beta,\Gamma)$, i.e. that 
\begin{equation}
     \frac{D_\Gamma}{|\bar{\Gamma}|} = \frac{D_\beta}{|\bar{\beta}|} \, .
\end{equation}
Such condition automatically provides a specific value of the free parameter $\Omega^2$, that is
\begin{equation}
    \Omega^2=\left|\frac{\bar{\Gamma}}{\bar{\beta}}\right| \, .
\end{equation}
Another option is to require that the symmetry involves $\beta$ and $\alpha$, rather than $\Gamma$.
Up to higher orders, imposing that the two relative fluctuations coincide, 
\begin{equation}
     \frac{D_\alpha}{|\bar{\alpha}|} \simeq \frac{D_\beta}{|\bar{\beta}|} \, ,
\end{equation}
yields the following value:
\begin{equation}
     \Omega^2\simeq\frac{\bar \alpha}{\gamma(\bar \alpha)\left|\bar{\beta}\right|}\,.
\end{equation}
It is easy to show that the two versions of $\Omega^2$ coincide in the regime of small velocities, as is clear from the definition of $\Gamma(\alpha)$.
Note that this does not follow from the dynamics, but corresponds to a specific preparation ansatz: the initial state is chosen such that relative quantum fluctuations are equally distributed between conjugate variables. 

Under this assumption, the kinematical parameter $\Omega^2$ is determined by the known quantities of the system.

\newpage

\bibliographystyle{apsrev4-1}
\bibliography{afuzzy}

\end{document}